\documentclass[aps,rmp,twocolumn]{revtex4-2}
\usepackage{hyperref}
\usepackage{graphicx}
\usepackage{url}
\usepackage{amsmath}
\usepackage{amsfonts}

\newcommand\rr{\mathbf{r}}

\begin{document}
\title{The virial theorem with periodic boundary conditions}
  
\author{A. C. Maggs} 
\affiliation{ {CNRS UMR7083, ESPCI Paris, PSL University,} {10 rue Vauquelin},
  {75005,} {Paris}, {France}}

\begin{abstract}
  The virial theorem relates averages of kinetic energy and forces in confined
  systems. It is widely used to relate stresses in molecular simulation as
  measured at a boundary and in the interior of a system. In periodic systems,
  the theorem must be modified in order to produce useful results. We formulate
  the theorem so that it is valid for both confined and for periodic systems. We
  cross-check our formulation in a study of a small system using both Monte
  Carlo and molecular dynamics simulations.
\end{abstract}
\maketitle

\subsection*{Introduction}

The virial theorem~\cite{Swenson1983} relates averages
of the kinetic energy to those of a function of the forces. For instance, for a
particle confined in one-dimension described by position \(x\) and momentum
\(p\)
\begin{equation}
  \left  \langle p^2/m  \right \rangle = \left \langle x dV/dx \right \rangle = -\langle x F \rangle
  \label{eq:oned}
\end{equation}
where the potential energy is \(V(x)\) and the force on the particle is
\(F(x)\).  The original derivation was dynamic, but we can also work from
statistical mechanics looking at the average of the Poisson bracket \([G,H]\),
where \(H\) is the Hamiltonian of the system and \(G=xp\).  With
\(H=p^2/2m+V(x)\).  The Poisson bracket is then
\begin{equation} [G,H] = p \frac{\partial H}{\partial p} - x \frac{\partial
    H}{\partial x} =\frac{p^2}{m } - x \frac{\partial V}{\partial x}
\end{equation}
and we recognize the two contributions to eq.~(\ref{eq:oned}). We choose to work
in the canonical ensemble, appropriate for a molecular system coupled to a
thermostat so that the statistical weight of a configuration is given by
$e^{-\beta H}$. Let us consider, firstly, the kinetic contribution to the
identity, \(p^2/m\), using integration by parts.
\begin{equation}
  \begin{aligned}
    \left \langle p \frac{\partial H}{\partial p} \right \rangle &=
    \frac{1}{Z_p} \int_{-\infty}^{\infty} p \frac{\partial H}{\partial p}
    e^{-\beta H}\, dp \\&= -\frac{T}{Z_p} \int_{-\infty}^{\infty} dp\, p
    \frac{\partial e^{-\beta H}}{\partial p} =T
  \end{aligned}
\end{equation}
We have introduced the temperature \(T=\beta^{-1}\) and \(Z_p\) is the kinetic
contribution to the partition function. An identical transformation applies to
the spatial derivative in the identity, \(x dV/dx\), when the physical system is
confined by the potential energy for \(x \rightarrow \pm \infty\) so that
integration by parts does not introduce a boundary contribution.

In a dilute gas identities such as eq.~(\ref{eq:oned}) find immediate
application. There is a balance between the kinetic contribution,
\( \langle p^2/m\rangle \), and that due to the potential,
\(\langle x F\rangle \) which describes the interaction of atoms with confining
walls. We now see an immediate problem with the application of the virial
theorem in periodic systems: In the dilute-gas limit, there is no possibility of
balancing the kinetic contribution with the wall collisions. Working in either
toroidal space, or using replicated copies of a primary cell the classic
formulation, assuming unbounded motion, requires modification. We note that the
opposite limit of a dense, low temperature crystal was analyzed
in~\cite{Louwerse2006}, where it was concluded that the calculation of the
pressure requires a mathematical formulation which avoids entirely the classical
virial theorem. This paper contains a strong criticism of formulations with
replicated copies, which can mis-count contributions. Recent work and codes
avoid the use of the virial theorem in periodic simulations, and uses
alternative routes to the pressure~\cite{Thompson2009}.

In this paper we present a unified picture of the virial theorem valid for
confined and periodic boundary conditions.  We find extra boundary contributions
(coming from integration by parts of the spatial derivative) that must be added
to the classical result. These contributions, for instance, allow one to
describe the dilute gas limit without contradiction. This formulation leads to a
transparent formulation of the virial route to the pressure, linking the
momentum transfer at the boundary of a simulation cell to the average volume
virial.

In a second part we test our expressions in high statistics simulations with
Monte Carlo and molecular dynamics studying a system of four particles in two
dimensions. We find that the constraint of conservation of momentum in molecular
dynamics leads to a non-uniform one-particle density that we study in detail.

\subsection*{Periodic formulation}

We consider \(N\) particles, in a \(d\)-dimensional periodic space. We allow the
particles to interact with a short ranged, central pair-potential
\(v(\rr_i -\rr_j )\) as well as an external one-body potential \(\phi(\rr_i) \)
so that
\begin{equation}
  V = \sum_{i<j} v(\rr_i -\rr_j ) + \sum_i \phi(\rr_i) \end{equation}
We only consider systems in their toroidal
representation so that \( 0\le r_{i\alpha} <L_\alpha\), with \(L_\alpha\) the box dimension in
direction \(\alpha\). The particles interact with a ``minimum image'' convention, so that the force
is not always parallel to the vector \(\rr_i-\rr_j\), but can rather be parallel to the vector
\begin{equation}
  \Delta_{ij}=\rr_i-\rr_j - \textbf{L}_n \label{eq:delta}
\end{equation}
where the vector \(\textbf{L}_n\) has entries \(n_\alpha L_\alpha\) with
\(n_\alpha \in \{-1,0,1\}\). We consider short-range, smooth potentials so that
interactions occur with at most a single value of the vector \(\textbf{L}_n\).
Use of the external potential \(\phi\) will enable us to continuously pass from
the case of an unconstrained periodic system with \(\phi=0\), to a system with
strong confining walls, so that we can compare the virial theorem in the two
limits.

We introduce the partial virial \(G_x = \sum_{i} x_i p_i\), with \(p_i\) the
$x$-component of the momentum for particle \(i\). As in the one-dimensional case
presented in the introduction, the treatment of the kinetic contribution to
\(\langle [G_x,H]\rangle \) is elementary
\begin{equation}
  \sum^N_{i} \left \langle p_i \frac{\partial H}{\partial p_i} \right \rangle = NT
\end{equation}
More care is required for the treatment of the configurational average. Consider
the term \(x_1F^x_{1}\), with \(F^x_1 \) the $x$-component of the force on
particle $1$: the partition function is defined as an integral over a finite
interval of \( x_1\). When performing the integration by parts, one must be
careful not drop the boundary contribution.
\begin{equation}
  \left \langle x_1 \frac{\partial V}{\partial x_1} \right \rangle = -\frac{T}{Z_r} \int_0^{L_x}
  d^{N}\rr\, x_1 \frac{\partial}{\partial x_1} e^{-\beta V}
\end{equation}
\(Z_r\) is the spatial contribution to the partition function.  We find two
contributions
\begin{equation}
  \begin{aligned}
    & -\frac{T}{Z_r} \left ( \int_0^{L_x} d^{N-1}\rr\, {[ x_1 e^{-\beta V}
        ]}_{x_1=0}^{x_1=L_x} - \int_0^{L_x} d^N\rr\, e^{-\beta V} \right ) \\&=
    - TL_x\bar \rho_1(x_1=L_x) +T
  \end{aligned}
\end{equation}
where $\bar \rho_1(x_1=0)\equiv \bar \rho_1(x_1=L_x)$ is the one-particle
density, \(\rho(\rr) = \langle \delta(\rr-\rr_1) \rangle \), integrated over the
boundary \(x=L_x\) of the fundamental cell.

Thus, the full periodic version of the virial theorem in the $x$-direction is
\begin{equation}
  \sum_{i} \left  \langle \frac{{p_i}^2}{m}  +  x_i F^x_i \right \rangle =   T L_x \bar \rho_N(x=L_x)
  \label{eq:th}
\end{equation}
with \(\bar \rho_N(x=L_x)\) the integrated density of the \(N\) particles over
the wall at \(x=L_x\).  The limit of weak potentials is now reasonable because
then $\bar \rho_N(x=L_x)= N/L_x$ and the theorem now allows finite kinetic
energy in the limit \(V\rightarrow 0\): We find the clearly correct result
\(\sum_i p_i^2/m = NT\)

For periodic systems, with \(\phi=0\) but with \(v(\rr)\) not zero, the average
of the kinetic energy and the boundary contribution cancel, and we find that the
average configurational virial vanishes.
\begin{equation}
  \sum_{i} {\left \langle  x_i \frac{\partial    V}{\partial x_i } \right \rangle}_{\phi=0} = 0 \label{eq:vv}
\end{equation}
However, when we impose a large positive potential \(\phi\) at the boundary of
the simulation cell then \( \bar \rho_N(x=L_x) \) is small and the result
eq.~(\ref{eq:th}) reduces to the confined limit, eq.~(\ref{eq:oned}).
Eq.~(\ref{eq:vv}) is compatible with the statements in~\cite{Louwerse2006} where
a perfect crystal at zero temperature is analyzed. The authors show that use of
eq.~(\ref{eq:vv}) in the wrong context leads to the conclusion that such crystal
is always at pressure $P=0$.

\subsection*{Link to Pressure}

\begin{figure}[ht]
  \includegraphics[width=0.5 \textwidth]{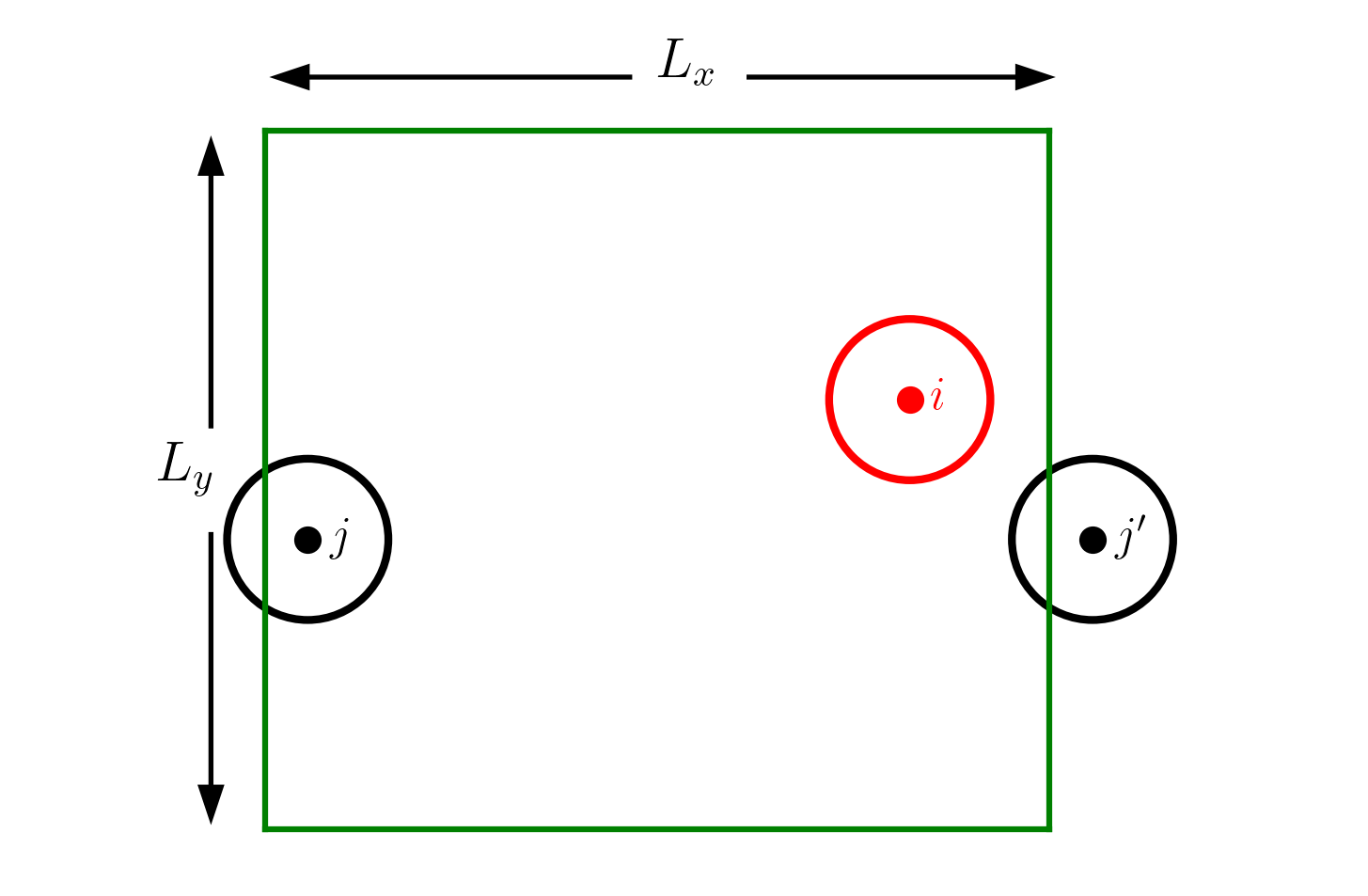}
  \caption{Two particles, $i$, $j$ have their centers in the periodic cell of
    volume $\Omega=L_x\times L_y$. The force between the particles is parallel
    to the vector joining $i$ and the nearest image of $j$, $j'$,
    eq.~(\ref{eq:delta}). The internal virial is calculated between $i$ and
    $j'$, requiring a balancing contribution of $L_x f^x_{ij}$ to the second
    line of eq.~(\ref{eq:kirk2}) which is to be interpreted as a contribution to
    the momentum flux at the cell edge $x=L_x$. Here
    \(n_x=1\).}\label{fig:flux}\end{figure}

It is conventional to re-write the potential part of the virial for pair
potentials in a form which is independent of the origin of the system. We do
this by breaking up the total force on particle \(i\) due to the potential $v$
in the following manner.  \( f^x_i = \sum_j f^x_{ij} \) where \(f^x_{ij}\)
denotes the $x$-component of the pair force on particle \(i\) due to \(j\), and
\(f^x_i\) is the $x$-component of the total pair force. Then,
\begin{equation}
  \sum^N_i x_i f^x_i = \sum^N_{i\ne j} x_i f^x_{ij}
\end{equation}
Note it is always true that \(f^x_{ij} = -f^x_{ji}\) even if,
Fig.~(\ref{fig:flux}), the interactions occur with images outside the primitive
cell so that
\begin{equation}
  \sum^N_i x_i f^x_i =\frac{1}{2} \sum^N_{i\ne j} (x_i- x_j) f^x_{ij}
  \label{eq:primitive}
\end{equation}
We now re-write eq.~(\ref{eq:primitive}) in terms of nearest image interactions
using \(\Delta_{ij}\), eq.~(\ref{eq:delta}), together with a boundary term and
find
\begin{equation}
  \sum^N_i x_i f^x_i =\frac{1}{2} \sum^N_{i\ne j} (\Delta_{ij,x} f^x_{ij} + n_x L_{x} f^x_{ij})
  \label{eq:image}
\end{equation}
We thus write eq.~(\ref{eq:th}) as
\begin{align}
  NT &+ \frac{1}{2} \sum^N_{i\ne j} \left \langle \Delta_{ij} f^x_{ij} \right \rangle - \sum^N_i
       \langle x_i \frac{\partial \phi} {\partial x_i}  \rangle  \label{eq:kirk1}
  \\ &= TL_x \bar \rho_N(L_x)
       -L_x \left \langle
       \sum_{\parbox{3em}{\tiny boundary pairs}}
       f^x_{ij} n_x \right \rangle \label{eq:kirk2}
  \\&=\Omega P_x 
\end{align}
We now interpret the contributions to
eq.~(\ref{eq:kirk1},~\ref{eq:kirk2}). Eq.~(\ref{eq:kirk1}) contains the usual
Irving-Kirkwood~\cite{Irving1950} stress tensor evaluated with a nearest image
convention together with a body force imposed by $\phi$.  Eq.~(\ref{eq:kirk2})
corresponds to momentum flux through the boundary $(x=L_x)$, giving the
$x$-component of the pressure \(P_x\) in a system of volume \(\Omega\). There
are two contributions to this momentum flux, a part due to particles crossing
the boundary, \(T \bar \rho_N(L_x) \), as well as a contribution involving
forces between particles which are on two different sides of the boundary in the
nearest image convention, Fig.~(\ref{fig:flux}).  The combination
\(f^x_{ij}n_x\) with now \(n_x \in \{-1,1\}\) orients the force through the
boundary, independent of the labeling order \(ij\).  In Fig.~(\ref{fig:flux}),
for the rightmost boundary, when \(i\) is inside the cell, and \(j'\) outside
\(n_x=1\). The position of the boundary in our derivation is
  arbitrary, so that eq.~\ref{eq:kirk2}) holds for an arbitrary choice of origin.

Again, we see that including the boundary term, arising from integration by
parts is essential for the correct link between the volume expression for the
stress tensor, and the momentum flux through the boundary of the simulation
cell, and thus the $x$-component of the pressure. Note the stress tensor,
\(\sigma\) and thus the pressure is not constant in the presence of a general
one-body potential~\cite{Davis19822}, as is implied by the Yvon-Born-Green
equation linking one-particle and two-particle correlations.

\subsection*{Dynamic approaches}
The derivation of the virial in periodic systems often works from a dynamic
approach, rather than the statistical approach used
here~\cite{Erpenbeck1984,Tsai1979}. This formulation follows the particles
originating in one copy of the simulation cell, now in the micro-canonical
ensemble obeying Newton's equation of motion, without folding the particles back
when they leave the fundamental copy. One looks at the time derivative of
\(G_x\) using the equations of motion, and then argues that since motions and
energies are bounded
\begin{equation}
  \lim_{t\to\infty }\frac{1}{t} \int_0^t \dot G_x(\tau)\, d\tau = 0
  \label{eq:av}
\end{equation}
If we follow the trajectories of particles, without folding them back into the
primary simulation domain, we find (see for instance~\cite{Erpenbeck1984,
  Winkler1992}) a result similar to eq.~(\ref{eq:kirk2}), with however important
differences. Firstly, the boundary contribution in \(T\bar \rho_N\) is entirely
missing, secondly the contribution \( L_n f^x_{ij} \) requires the uses of
arbitrary large integers \(n_\alpha \in \mathbb{Z} \) to push particles that
have moved large distances back into the primary cell in order to calculate the
system energy. In this limit it is impossible to recognize the result as
expressing the momentum flux evaluated at the boundary of the simulation cell.

We now show that by confining particles to the primary cell, and periodically
``transporting'' them when they cross the cell boundary we also find the same
formulation of the virial theorem eq.~(\ref{eq:th}).  We write
\begin{equation}
  \frac{d G_x}{dt} = \sum_i (\dot x_i p_i + x_i \dot p_i) = \sum_i  \left (\frac{p_i^2}{m} + x_i
    F^x_i \right )
  \label{eq:dyn}
\end{equation}
for time intervals where no particle leaves the primary domain. We define the
temperature from the average kinetic energy, so that
\( \langle p_i^2/m \rangle = T\). For a system that undergoes molecular dynamics
evolution with bounded forces then there is also a discontinuity in \(x_i p_i\)
each time a particle leaves the simulation cell, for instance at \(x=L_x\) and
is transported to \(x=0\). We thus need to add to eq.~(\ref{eq:dyn}) the extra
contribution \( -p_i L_x \delta(t-t_r) \) where $t_r$ is the time the particle
leaves to the right. A similar term is present when the particles leave to the
left. Taking the time average of eq.~(\ref{eq:dyn}) and noting that the
left-hand side averages to zero, eq.~(\ref{eq:av})
\begin{equation}
  \sum_i
  \left \langle \frac{p_i^2}{m} + x_i F^x_i \right \rangle =\frac{L_x}{t} \left [ \sum_{t_r} p_r -
    \sum_{t_l} p_l \right ] \label{eq:surface}
\end{equation}
where \(p_r\) and \(p_l\) is the momentum transported leaving the cell on the
right or left.  \(p_r\) and \(-p_l\) are both positive.  The sums on the right
are over all events where a particle leaves the primary domain. We now note this
momentum flux is just the perfect gas stress contribution \( T \bar \rho_N \) to
the pressure, and we find an expression identical to eq.~(\ref{eq:th}).

\subsection*{Numerical tests with 4 particles}
\begin{figure}[ht]
  \includegraphics[width=0.35\textwidth]{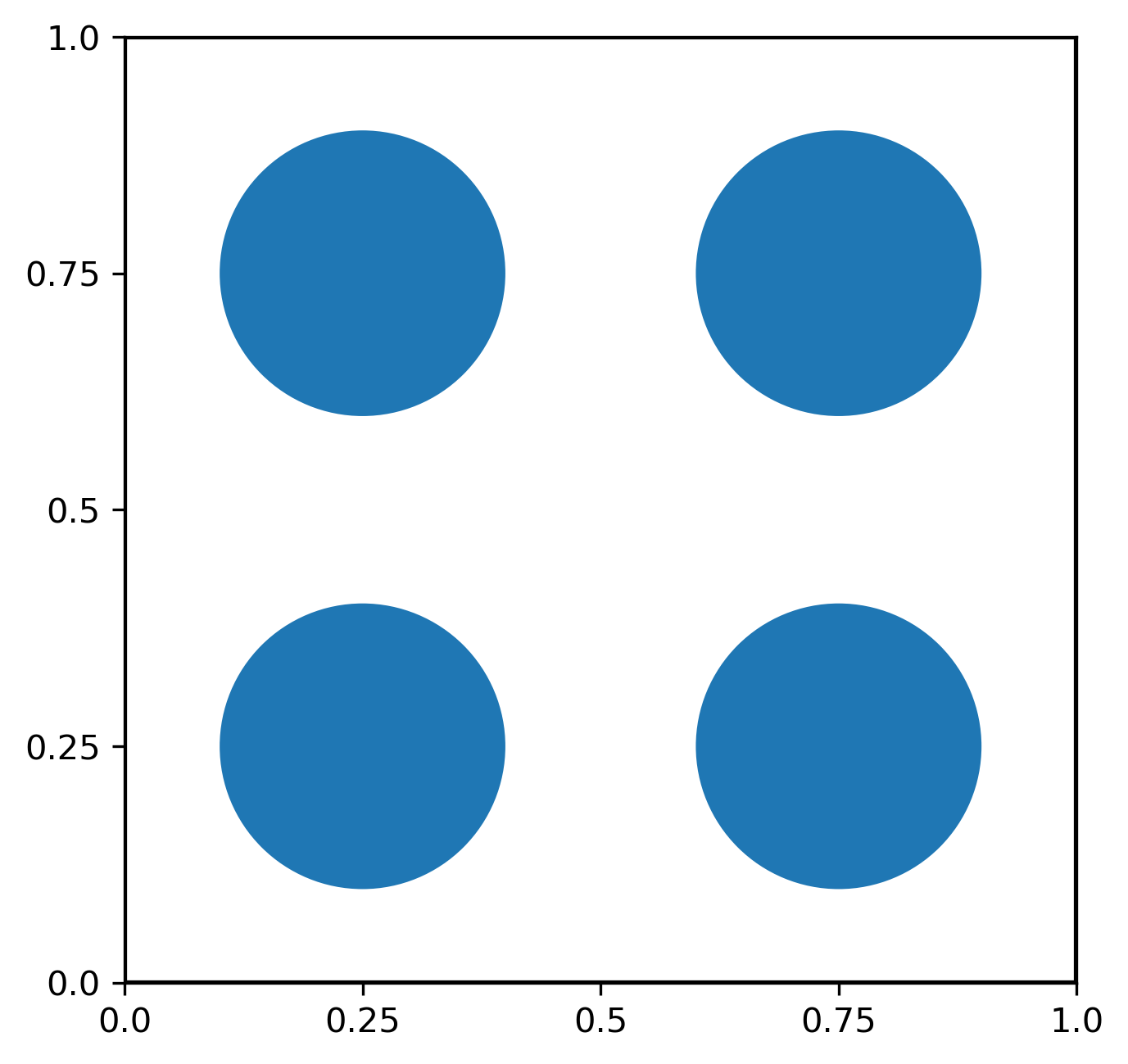}
  \caption{Four particles of radius $\sigma=0.15$ are simulated in a square
    periodic cell with $L_x=L_y=1$.}\label{fig:start}\end{figure}

We now consider in more detail the limit of a small number of
particles~\cite{Hoover1967,Ray1999,Wood2000}, and show detailed calculations for
$N=4$ with hard disks where efficient event driven methods are used to generate
high statistics data~\cite{Michel2014,Li2022}. These simulations allow us to
validate the expressions that we have found from the virial theorem. We first
simulate with event-chain Monte Carlo to find a reference value of the
pressure. For particles of radius $\sigma=0.15$ in a unit periodic cell,
Fig.~(\ref{fig:start}), we find \( \beta PV = 7.120 986(9) \). We separate the
pressure into a perfect gas contribution, $\beta V P^{perf} =4$, and a virial
contribution \( \beta V P^{vir} = 3.120 986(9) \).

In our molecular dynamics simulations we randomly initialize the velocities, and
remove the center of mass motion. We then scale velocities of particles so that
\begin{equation}
  \sum_i \frac{\textbf{v}_i^2}{2} = T (N-1) \label{eq:T}
\end{equation}
where we take the particles masses, $m=1$.
We note that the perfect gas limit of the pressure in molecular dynamics with
periodic boundary conditions is
\begin{equation}
  P = \frac{T}{V}(N-1) = T \rho \frac{(N-1)}{N}\label{eq:perfect}
\end{equation}
since we work in the center of mass frame of the system. In this we diverge from
historic papers on the small $N$ limit, where the temperature in molecular
dynamics is defined from $PV=NT$ for a dilute gas.

There are two routes available to evaluate the pressure in molecular
dynamics. The first, eq.~(\ref{eq:kirk2}), is the sum of the momentum transfer
from collisions across the cell boundary (for instance a collision between $i$
and $j'$ in Fig.~(\ref{fig:flux})), plus the momentum flux due to particles
travelling through the boundary which we evaluate with the right-hand side of
eq.~(\ref{eq:surface}).  We find $\beta P V= 3 + 3.120 995(9) $. Where we have
again separated the perfect gas eq.~(\ref{eq:perfect}) and virial
contributions. The virial contribution is identical, within statistical errors,
to that found given by Monte Carlo methods. We also checked that the
contribution to the momentum flux due to particles crossing the boundary
corresponds to the term $T \bar \rho_N(L_x) $ in eq.~(\ref{eq:kirk2}) with the
modification of eq.~(\ref{eq:perfect}). The transport momentum flux is thus
$T \bar \rho_N(L_x) (N-1)/N $.

\begin{figure}[ht]
  \includegraphics[width=0.4\textwidth]{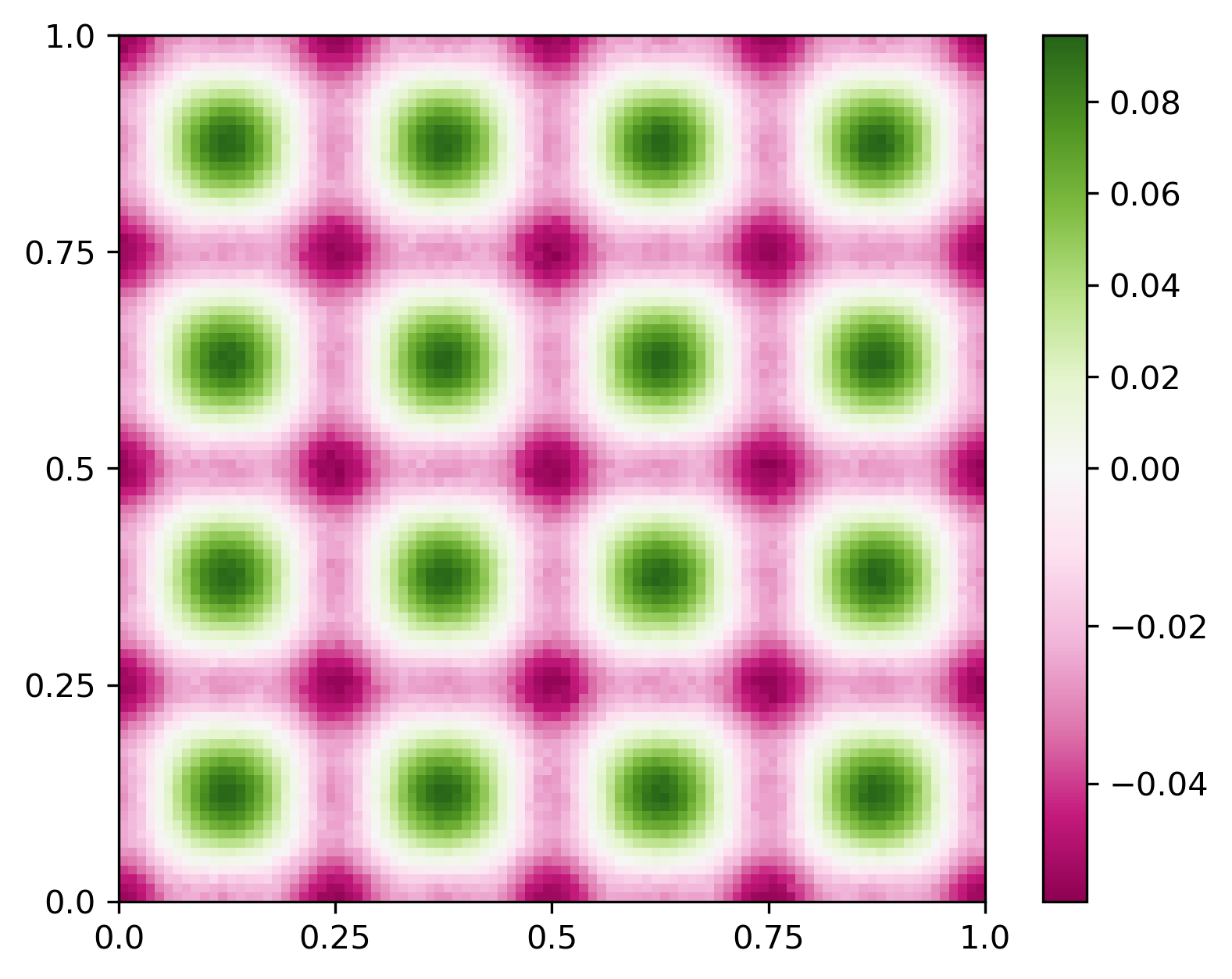}
  \caption{Histogram of particle positions from molecular dynamics simulation of
    Fig.~(\ref{fig:start}), $\sigma=0.15$.  We find a regular $4 \times 4$
    pattern of peaks in the density. Local density varies by $ \sim 8\%$ from
    the mean.  The particle positions of Fig.~(\ref{fig:start}) correspond to a
    minimum in the probability density.}\label{fig:pattern}\end{figure}

When examining our simulations,  we found that although the sum of the two
contributions to the pressure is statistically constant throughout the
simulation cell, the magnitude of each contribution depends on the initial
conditions of the simulation and is inhomogeneous in space.  We investigated the
variation in the two contributions to the pressure in detail and found,
(Fig.~(\ref{fig:pattern})), that a histogram of particle positions is
non-uniform. When working with $N$ particles we find a grid of $N \times N$
peaks in the density. We interpret this inhomogeneity as being due to the
conservation of the position of the center of mass in molecular dynamics
simulation: When a single particle moves to the right by a distance
$L_x (1-1/N)$ and all other particles move to the left by a distance $L_x/N$ we
generate an equivalent configuration with the same center of mass, that appears
shifted left by $L_x/N$. Each peak in the grid of Fig.~(\ref{fig:pattern})
corresponds to configurations with the same relative positions of all
particles. Depending on the position of the center of mass with respect to the
cell boundaries at the start of the simulation, the two contributions to the
momentum flux vary while their sum remains identical.

The second route to the pressure in molecular dynamics is through the time
averaged virial eq.~\ref{eq:kirk1}, with again the perfect gas expression $NT$,
replaced with $(N-1)T$. In our molecular dynamics simulations the virial and
momentum routes to the pressure agree to one part in $10^{10}$, confirming the
correctness of our derivation of the periodic theorem, and its application in a
system with non-trivial spatial structure.

\begin{figure}[ht]
  \includegraphics[width=0.4\textwidth]{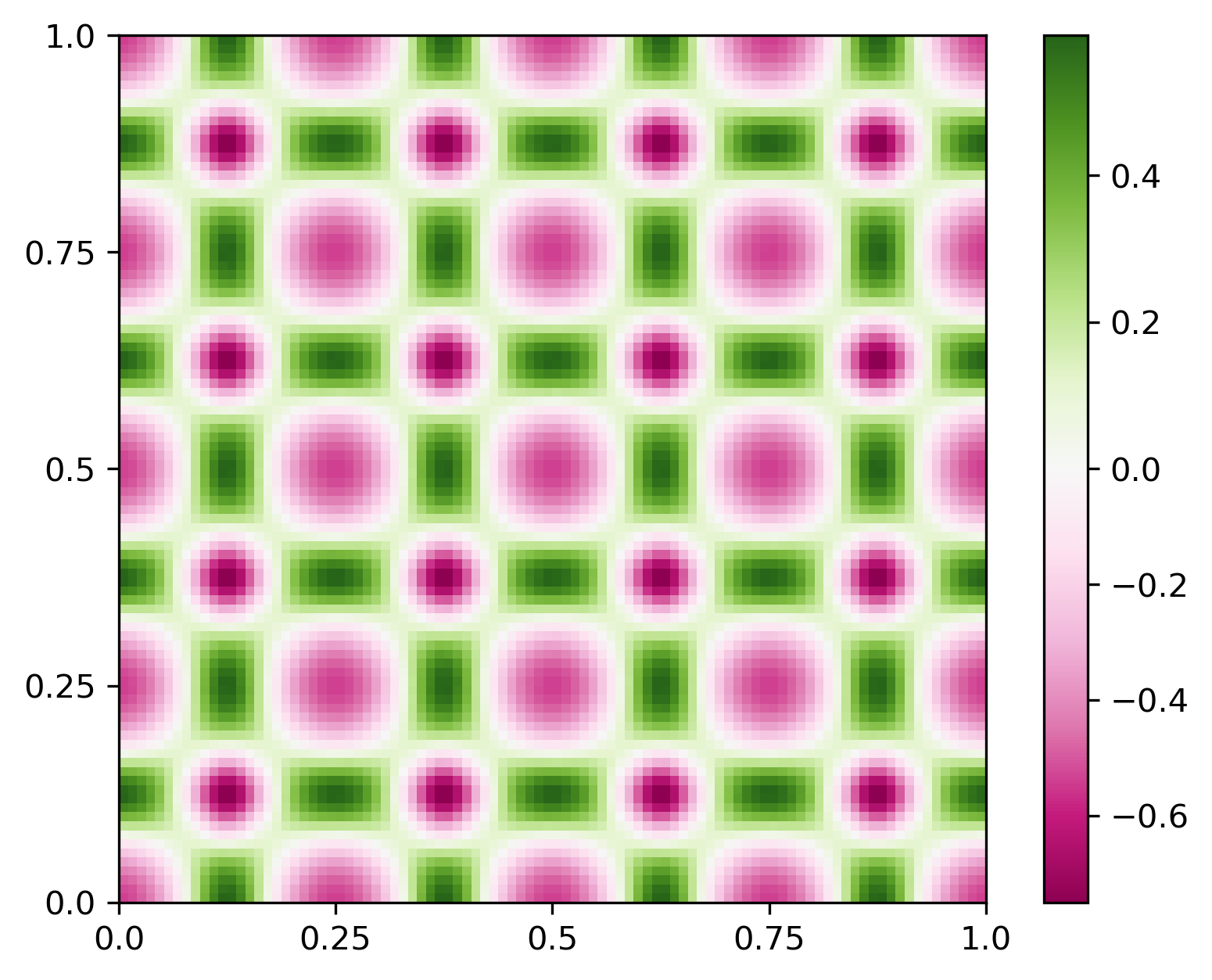}
  \caption{Density variations from a molecular dynamics simulation with 4
    particles with $\sigma=0.22$. Positions that are a maximum of density at
    smaller particle size here become a deep density
    minimum.}\label{fig:pattern2}\end{figure}

We are able to follow in our binning procedure the pattern of $N^2$ peaks up to
$N=6$ (working at constant volume fraction of particles), the amplitude of the
pattern becomes rapidly smaller with increasing $N$. We also confirmed the
structure for $N=7$ with Fourier analysis. We were also able to study the
pattern of peaks in our event-chain Monte Carlo code by measuring the density in
a reference frame moving with the center of mass, then folding positions back
into a co-moving simulation cell.  Non-trivial density patterns occur in Monte
Carlo when measured in the correct reference frame. Increasing the particle size
leads a to a more complicated pattern, Fig.~(\ref{fig:pattern2}) of density
variations.

\subsection*{Linking kinetic energy and temperature}
We now demonstrate, under the usual hypothesis of factorization of velocity and
spatial degrees of freedom, that the kinetic virial from molecular dynamics and
the thermodynamic virial measured in Monte Carlo simulation are identical when
using the correct link between temperature and kinetic energy,
eq.~(\ref{eq:T}). We use the approach of~\cite{Li2022}, section~III.c.2, where
the pressure is calculated in a thermodynamic approach.  If $\hat g(r)$ is the
probability density for finding a given pair of particles near separation $r$,
then the contribution to the pressure coming from pair interactions is
\begin{equation}
  \beta P^{vir} =  \frac{N(N-1)}{V} 2 \pi  \sigma^2 \hat g(2 \sigma)
  \label{eq:pair}
\end{equation}

The kinetic virial as measured in a molecular dynamics simulation for hard disks
is given by
\begin{equation}
  P^{vir} =  - \frac{1}{2 V} \left  \langle  (\textbf{v}_i-\textbf{v}_j) \cdot
    (\textbf{r}_i-\textbf{r}_j) \right  \rangle
\end{equation}
where the expectation is a time average of pair collisions between pairs of
particles $i$ and $j$.  This time average is calculated by averaging over the
flux of collisions, $f$:
\begin{equation}
  f=\frac{N(N-1)}{2} 2 \pi (2 \sigma) \hat g(2 \sigma) \frac{\big [{v^r}_1-{v^r}_2
    \big ] }{2} H(\textbf{v})
\end{equation} 
Where $H(\textbf{v})$ is a uniform distribution of $2N$ velocity components on
the hypersphere, confined to $\sum_i \textbf{v}_i =0$ and
$ \big [{v^r}_1-{v^r}_2 \big ] $ is the radial component of the relative
velocity of two particles.  The contribution to the pressure is thus
\begin{equation}
  P^{vir} =    \frac{N(N-1)}{4 V} 2 \pi {(2 \sigma)}^2 \hat g(2 \sigma) \left \langle
    \frac {{\big  [      {v^r}_1-{v^r}_2 \big ]}^2}{2}  H(\textbf{v} ) \right \rangle
\end{equation}
The velocity average is calculated by noting that
$ \left \langle {v^r}_1{v^r}_2 H(\textbf{v})\right \rangle = - \langle {
  ({v^r}_1 )}^2 H(\textbf{v})\rangle /(N-1)$ so that
\begin{equation}
  P^{vir} =     \frac{N(N-1)}{4V} 2 \pi {(2 \sigma)}^2 {\hat g} (2\sigma)
  \left \langle  {(v^r_1)} ^2 H(\textbf{v}) \right \rangle \frac{N}{N-1} \label{eq:finalmd}
\end{equation}
Consistency between eqs.~(\ref{eq:finalmd},~\ref{eq:pair}) requires
\begin{equation}
  \left \langle \frac{\textbf{v}_i^2}{2} \right  \rangle = \frac{N-1}{N} T
\end{equation}
as used in our simulations.

\subsection*{Conclusion}

We have treated the generalization of the virial theorem to periodic systems
using two approaches. Firstly, using ensemble averages from equilibrium
statistical mechanics, secondly using a dynamical approach from the equations of
motion. Our approach avoids the difficulties carefully pointed out
in~\cite{Louwerse2006}, which lead to nonsensical results for the pressure, when
starting from classical statements of the theorem.

Our treatment of the virial theorem has features that are similar to the quantum
virial theorem treated in~\cite{Abad1991,Esteve2012}. Again, the use of a
periodic space in quantum mechanics leads to extra end-corrections to the
theorem, here involving boundary values of the wavefunction. Our treatment deals
cleanly with the weakly interacting gas limit, which is clearly not correctly
given using the classical virial expression. Using an external one-body
potential, we interpolate between the standard confined simulation box, and a
system where particles move freely through cell boundaries. This one-body
potential can describe either a high uniform wall, or more complicated
geometries such as a hole in an otherwise impenetrable barrier.  Our expression
gives a direct link between the Irving-Kirkwood expression for the stress tensor
and the boundary momentum flux.  We checked the use of our virial and momentum
estimators for the case of $N=4$ particles, and compared the results to
independent event-chain Monte Carlo simulations.

Open questions remain for the formulation of the virial for long-ranged
electrostatic interactions. The most natural formulation of electrostatics
boundary conditions requires the use of a non-periodic representation of the
particle positions in order to correctly represent the molecular polarization of
charged media~\cite{Caillol1994,Maggs2004,Sprik2018,maggs2002,Maggs2012}. 

\subsection*{Acknowledgments} 
I would like to thank Werner Krauth for the many discussions that formed the starting point of this work.

\bibliography{virial}

\begin{thebibliography}{20}%
\makeatletter
\providecommand \@ifxundefined [1]{%
 \@ifx{#1\undefined}
}%
\providecommand \@ifnum [1]{%
 \ifnum #1\expandafter \@firstoftwo
 \else \expandafter \@secondoftwo
 \fi
}%
\providecommand \@ifx [1]{%
 \ifx #1\expandafter \@firstoftwo
 \else \expandafter \@secondoftwo
 \fi
}%
\providecommand \natexlab [1]{#1}%
\providecommand \enquote  [1]{``#1''}%
\providecommand \bibnamefont  [1]{#1}%
\providecommand \bibfnamefont [1]{#1}%
\providecommand \citenamefont [1]{#1}%
\providecommand \href@noop [0]{\@secondoftwo}%
\providecommand \href [0]{\begingroup \@sanitize@url \@href}%
\providecommand \@href[1]{\@@startlink{#1}\@@href}%
\providecommand \@@href[1]{\endgroup#1\@@endlink}%
\providecommand \@sanitize@url [0]{\catcode `\\12\catcode `\$12\catcode
  `\&12\catcode `\#12\catcode `\^12\catcode `\_12\catcode `\%12\relax}%
\providecommand \@@startlink[1]{}%
\providecommand \@@endlink[0]{}%
\providecommand \url  [0]{\begingroup\@sanitize@url \@url }%
\providecommand \@url [1]{\endgroup\@href {#1}{\urlprefix }}%
\providecommand \urlprefix  [0]{URL }%
\providecommand \Eprint [0]{\href }%
\providecommand \doibase [0]{https://doi.org/}%
\providecommand \selectlanguage [0]{\@gobble}%
\providecommand \bibinfo  [0]{\@secondoftwo}%
\providecommand \bibfield  [0]{\@secondoftwo}%
\providecommand \translation [1]{[#1]}%
\providecommand \BibitemOpen [0]{}%
\providecommand \bibitemStop [0]{}%
\providecommand \bibitemNoStop [0]{.\EOS\space}%
\providecommand \EOS [0]{\spacefactor3000\relax}%
\providecommand \BibitemShut  [1]{\csname bibitem#1\endcsname}%
\let\auto@bib@innerbib\@empty
\bibitem [{\citenamefont {Abad}\ and\ \citenamefont {Esteve}(1991)}]{Abad1991}%
  \BibitemOpen
  \bibfield  {author} {\bibinfo {author} {\bibnamefont {Abad}, \bibfnamefont
  {J}}, and\ \bibinfo {author} {\bibfnamefont {J.~G.}\ \bibnamefont {Esteve}}}
  (\bibinfo {year} {1991}),\ \bibfield  {title} {\enquote {\bibinfo {title}
  {Generalized virial theorem for compact problems},}\ }\href
  {https://doi.org/10.1103/PhysRevA.44.4728} {\bibfield  {journal} {\bibinfo
  {journal} {Phys. Rev. A}\ }\textbf {\bibinfo {volume} {44}},\ \bibinfo
  {pages} {4728--4729}}\BibitemShut {NoStop}%
\bibitem [{\citenamefont {Caillol}(1994)}]{Caillol1994}%
  \BibitemOpen
  \bibfield  {author} {\bibinfo {author} {\bibnamefont {Caillol}, \bibfnamefont
  {Jean‐Michel}}} (\bibinfo {year} {1994}),\ \bibfield  {title} {\enquote
  {\bibinfo {title} {Comments on the numerical simulations of electrolytes in
  periodic boundary conditions},}\ }\href {https://doi.org/10.1063/1.468422}
  {\bibfield  {journal} {\bibinfo  {journal} {The Journal of Chemical Physics}\
  }\textbf {\bibinfo {volume} {101}}~(\bibinfo {number} {7}),\ \bibinfo {pages}
  {6080--6090}}\BibitemShut {NoStop}%
\bibitem [{\citenamefont {Davis}\ and\ \citenamefont
  {Scriven}(1982)}]{Davis19822}%
  \BibitemOpen
  \bibfield  {author} {\bibinfo {author} {\bibnamefont {Davis}, \bibfnamefont
  {H~T}}, and\ \bibinfo {author} {\bibfnamefont {L.~E.}\ \bibnamefont
  {Scriven}}} (\bibinfo {year} {1982}),\ \enquote {\bibinfo {title} {Stress and
  structure in fluid interfaces},}\ in\ \href
  {https://doi.org/https://doi.org/10.1002/9780470142691.ch6} {\emph {\bibinfo
  {booktitle} {Advances in Chemical Physics}}}\ (\bibinfo  {publisher} {John
  Wiley \& Sons, Ltd})\ pp.\ \bibinfo {pages} {357--454}\BibitemShut {NoStop}%
\bibitem [{\citenamefont {Erpenbeck}\ and\ \citenamefont
  {Wood}(1984)}]{Erpenbeck1984}%
  \BibitemOpen
  \bibfield  {author} {\bibinfo {author} {\bibnamefont {Erpenbeck},
  \bibfnamefont {Jerome~J}}, and\ \bibinfo {author} {\bibfnamefont
  {William~W.}\ \bibnamefont {Wood}}} (\bibinfo {year} {1984}),\ \bibfield
  {title} {\enquote {\bibinfo {title} {Molecular dynamics calculations of the
  hard-sphere equation of state},}\ }\href {https://doi.org/10.1007/BF01014387}
  {\bibfield  {journal} {\bibinfo  {journal} {Journal of Statistical Physics}\
  }\textbf {\bibinfo {volume} {35}}~(\bibinfo {number} {3}),\ \bibinfo {pages}
  {321--340}}\BibitemShut {NoStop}%
\bibitem [{\citenamefont {Esteve}\ \emph {et~al.}(2012)\citenamefont {Esteve},
  \citenamefont {Falceto},\ and\ \citenamefont {Giri}}]{Esteve2012}%
  \BibitemOpen
  \bibfield  {author} {\bibinfo {author} {\bibnamefont {Esteve}, \bibfnamefont
  {J~G}}, \bibinfo {author} {\bibfnamefont {F.}~\bibnamefont {Falceto}}, and\
  \bibinfo {author} {\bibfnamefont {Pulak~Ranjan}\ \bibnamefont {Giri}}}
  (\bibinfo {year} {2012}),\ \bibfield  {title} {\enquote {\bibinfo {title}
  {Boundary contributions to the hypervirial theorem},}\ }\href
  {https://doi.org/10.1103/PhysRevA.85.022104} {\bibfield  {journal} {\bibinfo
  {journal} {Phys. Rev. A}\ }\textbf {\bibinfo {volume} {85}},\ \bibinfo
  {pages} {022104}}\BibitemShut {NoStop}%
\bibitem [{\citenamefont {Hoover}\ and\ \citenamefont
  {Alder}(1967)}]{Hoover1967}%
  \BibitemOpen
  \bibfield  {author} {\bibinfo {author} {\bibnamefont {Hoover}, \bibfnamefont
  {William~G}}, and\ \bibinfo {author} {\bibfnamefont {Berni~J.}\ \bibnamefont
  {Alder}}} (\bibinfo {year} {1967}),\ \bibfield  {title} {\enquote {\bibinfo
  {title} {Studies in molecular dynamics. iv. the pressure, collision rate, and
  their number dependence for hard disks},}\ }\href
  {https://doi.org/10.1063/1.1840726} {\bibfield  {journal} {\bibinfo
  {journal} {The Journal of Chemical Physics}\ }\textbf {\bibinfo {volume}
  {46}}~(\bibinfo {number} {2}),\ \bibinfo {pages} {686--691}}\BibitemShut
  {NoStop}%
\bibitem [{\citenamefont {Irving}\ and\ \citenamefont
  {Kirkwood}(1950)}]{Irving1950}%
  \BibitemOpen
  \bibfield  {author} {\bibinfo {author} {\bibnamefont {Irving}, \bibfnamefont
  {J~H}}, and\ \bibinfo {author} {\bibfnamefont {John~G.}\ \bibnamefont
  {Kirkwood}}} (\bibinfo {year} {1950}),\ \bibfield  {title} {\enquote
  {\bibinfo {title} {The statistical mechanical theory of transport processes.
  iv. the equations of hydrodynamics},}\ }\href
  {https://doi.org/10.1063/1.1747782} {\bibfield  {journal} {\bibinfo
  {journal} {The Journal of Chemical Physics}\ }\textbf {\bibinfo {volume}
  {18}}~(\bibinfo {number} {6}),\ \bibinfo {pages} {817--829}}\BibitemShut
  {NoStop}%
\bibitem [{\citenamefont {Li}\ \emph {et~al.}(2022)\citenamefont {Li},
  \citenamefont {Nishikawa}, \citenamefont {Höllmer}, \citenamefont {Carillo},
  \citenamefont {Maggs},\ and\ \citenamefont {Krauth}}]{Li2022}%
  \BibitemOpen
  \bibfield  {author} {\bibinfo {author} {\bibnamefont {Li}, \bibfnamefont
  {Botao}}, \bibinfo {author} {\bibfnamefont {Yoshihiko}\ \bibnamefont
  {Nishikawa}}, \bibinfo {author} {\bibfnamefont {Philipp}\ \bibnamefont
  {Höllmer}}, \bibinfo {author} {\bibfnamefont {Louis}\ \bibnamefont
  {Carillo}}, \bibinfo {author} {\bibfnamefont {A.~C.}\ \bibnamefont {Maggs}},
  and\ \bibinfo {author} {\bibfnamefont {Werner}\ \bibnamefont {Krauth}}}
  (\bibinfo {year} {2022}),\ \bibfield  {title} {\enquote {\bibinfo {title}
  {Hard-disk pressure computations—a historic perspective},}\ }\href
  {https://doi.org/10.1063/5.0126437} {\bibfield  {journal} {\bibinfo
  {journal} {The Journal of Chemical Physics}\ }\textbf {\bibinfo {volume}
  {157}}~(\bibinfo {number} {23}),\ \bibinfo {pages} {234111}}\BibitemShut
  {NoStop}%
\bibitem [{\citenamefont {Louwerse}\ and\ \citenamefont
  {Baerends}(2006)}]{Louwerse2006}%
  \BibitemOpen
  \bibfield  {author} {\bibinfo {author} {\bibnamefont {Louwerse},
  \bibfnamefont {Manuel~J}}, and\ \bibinfo {author} {\bibfnamefont {Evert~Jan}\
  \bibnamefont {Baerends}}} (\bibinfo {year} {2006}),\ \bibfield  {title}
  {\enquote {\bibinfo {title} {Calculation of pressure in case of periodic
  boundary conditions},}\ }\href
  {https://doi.org/https://doi.org/10.1016/j.cplett.2006.01.087} {\bibfield
  {journal} {\bibinfo  {journal} {Chemical Physics Letters}\ }\textbf {\bibinfo
  {volume} {421}}~(\bibinfo {number} {1}),\ \bibinfo {pages}
  {138--141}}\BibitemShut {NoStop}%
\bibitem [{\citenamefont {Maggs}(2002)}]{maggs2002}%
  \BibitemOpen
  \bibfield  {author} {\bibinfo {author} {\bibnamefont {Maggs}, \bibfnamefont
  {A~C}}} (\bibinfo {year} {2002}),\ \bibfield  {title} {\enquote {\bibinfo
  {title} {Dynamics of a local algorithm for simulating {C}oulomb
  interactions},}\ }\href {https://doi.org/10.1063/1.1487821} {\bibfield
  {journal} {\bibinfo  {journal} {The Journal of Chemical Physics}\ }\textbf
  {\bibinfo {volume} {117}}~(\bibinfo {number} {5}),\ \bibinfo {pages}
  {1975--1981}},\ \Eprint
  {https://arxiv.org/abs/https://doi.org/10.1063/1.1487821}
  {https://doi.org/10.1063/1.1487821} \BibitemShut {NoStop}%
\bibitem [{\citenamefont {Maggs}(2004)}]{Maggs2004}%
  \BibitemOpen
  \bibfield  {author} {\bibinfo {author} {\bibnamefont {Maggs}, \bibfnamefont
  {A~C}}} (\bibinfo {year} {2004}),\ \bibfield  {title} {\enquote {\bibinfo
  {title} {Auxiliary field {M}onte {C}arlo for charged particles},}\ }\href
  {https://doi.org/10.1063/1.1642587} {\bibfield  {journal} {\bibinfo
  {journal} {The Journal of Chemical Physics}\ }\textbf {\bibinfo {volume}
  {120}}~(\bibinfo {number} {7}),\ \bibinfo {pages} {3108--3118}}\BibitemShut
  {NoStop}%
\bibitem [{\citenamefont {Maggs}(2012)}]{Maggs2012}%
  \BibitemOpen
  \bibfield  {author} {\bibinfo {author} {\bibnamefont {Maggs}, \bibfnamefont
  {A~C}}} (\bibinfo {year} {2012}),\ \bibfield  {title} {\enquote {\bibinfo
  {title} {A minimizing principle for the {P}oisson-{B}oltzmann equation},}\
  }\href {https://doi.org/10.1209/0295-5075/98/16012} {\bibfield  {journal}
  {\bibinfo  {journal} {Europhysics Letters}\ }\textbf {\bibinfo {volume}
  {98}}~(\bibinfo {number} {1}),\ \bibinfo {pages} {16012}}\BibitemShut
  {NoStop}%
\bibitem [{\citenamefont {Michel}\ \emph {et~al.}(2014)\citenamefont {Michel},
  \citenamefont {Kapfer},\ and\ \citenamefont {Krauth}}]{Michel2014}%
  \BibitemOpen
  \bibfield  {author} {\bibinfo {author} {\bibnamefont {Michel}, \bibfnamefont
  {Manon}}, \bibinfo {author} {\bibfnamefont {Sebastian~C.}\ \bibnamefont
  {Kapfer}}, and\ \bibinfo {author} {\bibfnamefont {Werner}\ \bibnamefont
  {Krauth}}} (\bibinfo {year} {2014}),\ \bibfield  {title} {\enquote {\bibinfo
  {title} {Generalized event-chain monte carlo: Constructing rejection-free
  global-balance algorithms from infinitesimal steps},}\ }\href
  {https://doi.org/10.1063/1.4863991} {\bibfield  {journal} {\bibinfo
  {journal} {The Journal of Chemical Physics}\ }\textbf {\bibinfo {volume}
  {140}}~(\bibinfo {number} {5}),\ \bibinfo {pages} {054116}}\BibitemShut
  {NoStop}%
\bibitem [{\citenamefont {Ray}\ and\ \citenamefont {Zhang}(1999)}]{Ray1999}%
  \BibitemOpen
  \bibfield  {author} {\bibinfo {author} {\bibnamefont {Ray}, \bibfnamefont
  {John~R}}, and\ \bibinfo {author} {\bibfnamefont {Hongwei}\ \bibnamefont
  {Zhang}}} (\bibinfo {year} {1999}),\ \bibfield  {title} {\enquote {\bibinfo
  {title} {Correct microcanonical ensemble in molecular dynamics},}\ }\href
  {https://doi.org/10.1103/PhysRevE.59.4781} {\bibfield  {journal} {\bibinfo
  {journal} {Phys. Rev. E}\ }\textbf {\bibinfo {volume} {59}},\ \bibinfo
  {pages} {4781--4785}}\BibitemShut {NoStop}%
\bibitem [{\citenamefont {Sprik}(2018)}]{Sprik2018}%
  \BibitemOpen
  \bibfield  {author} {\bibinfo {author} {\bibnamefont {Sprik}, \bibfnamefont
  {M}}} (\bibinfo {year} {2018}),\ \bibfield  {title} {\enquote {\bibinfo
  {title} {Finite {M}axwell field and electric displacement {H}amiltonians
  derived from a current dependent {L}agrangian},}\ }\href
  {https://doi.org/10.1080/00268976.2018.1431406} {\bibfield  {journal}
  {\bibinfo  {journal} {Molecular Physics}\ }\textbf {\bibinfo {volume}
  {116}}~(\bibinfo {number} {21-22}),\ \bibinfo {pages}
  {3114--3120}}\BibitemShut {NoStop}%
\bibitem [{\citenamefont {Swenson}(1983)}]{Swenson1983}%
  \BibitemOpen
  \bibfield  {author} {\bibinfo {author} {\bibnamefont {Swenson}, \bibfnamefont
  {Robert~J}}} (\bibinfo {year} {1983}),\ \bibfield  {title} {\enquote
  {\bibinfo {title} {Comments on virial theorems for bounded systems},}\ }\href
  {https://doi.org/10.1119/1.13390} {\bibfield  {journal} {\bibinfo  {journal}
  {American Journal of Physics}\ }\textbf {\bibinfo {volume} {51}}~(\bibinfo
  {number} {10}),\ \bibinfo {pages} {940--942}}\BibitemShut {NoStop}%
\bibitem [{\citenamefont {Thompson}\ \emph {et~al.}(2009)\citenamefont
  {Thompson}, \citenamefont {Plimpton},\ and\ \citenamefont
  {Mattson}}]{Thompson2009}%
  \BibitemOpen
  \bibfield  {author} {\bibinfo {author} {\bibnamefont {Thompson},
  \bibfnamefont {Aidan~P}}, \bibinfo {author} {\bibfnamefont {Steven~J.}\
  \bibnamefont {Plimpton}}, and\ \bibinfo {author} {\bibfnamefont {William}\
  \bibnamefont {Mattson}}} (\bibinfo {year} {2009}),\ \bibfield  {title}
  {\enquote {\bibinfo {title} {General formulation of pressure and stress
  tensor for arbitrary many-body interaction potentials under periodic boundary
  conditions},}\ }\href {https://doi.org/10.1063/1.3245303} {\bibfield
  {journal} {\bibinfo  {journal} {The Journal of Chemical Physics}\ }\textbf
  {\bibinfo {volume} {131}}~(\bibinfo {number} {15}),\ \bibinfo {pages}
  {154107}}\BibitemShut {NoStop}%
\bibitem [{\citenamefont {Tsai}(1979)}]{Tsai1979}%
  \BibitemOpen
  \bibfield  {author} {\bibinfo {author} {\bibnamefont {Tsai}, \bibfnamefont
  {D~H}}} (\bibinfo {year} {1979}),\ \bibfield  {title} {\enquote {\bibinfo
  {title} {The virial theorem and stress calculation in molecular dynamics},}\
  }\href {https://doi.org/10.1063/1.437577} {\bibfield  {journal} {\bibinfo
  {journal} {The Journal of Chemical Physics}\ }\textbf {\bibinfo {volume}
  {70}}~(\bibinfo {number} {3}),\ \bibinfo {pages} {1375--1382}}\BibitemShut
  {NoStop}%
\bibitem [{\citenamefont {Winkler}\ \emph {et~al.}(1992)\citenamefont
  {Winkler}, \citenamefont {Morawitz},\ and\ \citenamefont
  {Yoon}}]{Winkler1992}%
  \BibitemOpen
  \bibfield  {author} {\bibinfo {author} {\bibnamefont {Winkler}, \bibfnamefont
  {RG}}, \bibinfo {author} {\bibfnamefont {H.}~\bibnamefont {Morawitz}}, and\
  \bibinfo {author} {\bibfnamefont {D.Y.}\ \bibnamefont {Yoon}}} (\bibinfo
  {year} {1992}),\ \bibfield  {title} {\enquote {\bibinfo {title} {Novel
  molecular dynamics simulations at constant pressure},}\ }\href
  {https://doi.org/10.1080/00268979200100491} {\bibfield  {journal} {\bibinfo
  {journal} {Molecular Physics}\ }\textbf {\bibinfo {volume} {75}}~(\bibinfo
  {number} {3}),\ \bibinfo {pages} {669--688}}\BibitemShut {NoStop}%
\bibitem [{\citenamefont {Wood}\ \emph {et~al.}(2000)\citenamefont {Wood},
  \citenamefont {Erpenbeck}, \citenamefont {Baker},\ and\ \citenamefont
  {Johnson}}]{Wood2000}%
  \BibitemOpen
  \bibfield  {author} {\bibinfo {author} {\bibnamefont {Wood}, \bibfnamefont
  {William~W}}, \bibinfo {author} {\bibfnamefont {Jerome~J.}\ \bibnamefont
  {Erpenbeck}}, \bibinfo {author} {\bibfnamefont {George~A.}\ \bibnamefont
  {Baker}}, and\ \bibinfo {author} {\bibfnamefont {J.~D.}\ \bibnamefont
  {Johnson}}} (\bibinfo {year} {2000}),\ \bibfield  {title} {\enquote {\bibinfo
  {title} {Molecular dynamics ensemble, equation of state, and ergodicity},}\
  }\href {https://doi.org/10.1103/PhysRevE.63.011106} {\bibfield  {journal}
  {\bibinfo  {journal} {Phys. Rev. E}\ }\textbf {\bibinfo {volume} {63}},\
  \bibinfo {pages} {011106}}\BibitemShut {NoStop}%
\end{thebibliography}%

\end{document}